\documentclass[12pt]{article}
\usepackage{amsmath,amssymb,amsthm,latexsym}
\usepackage{stmaryrd,wasysym,upgreek,mathrsfs,dsfont}
\usepackage[english]{babel}
\usepackage{graphicx,color}
\usepackage{amscd}

\usepackage[pdftex]{hyperref}
\newtheorem{lemma}{Lemma}[section]

\newtheorem{conjecture}{Conjecture}[section]

\newtheorem{definition}{Definition}[section]
\newtheorem{theorem}{Theorem}[section]
\newcommand{\be}{\begin{equation}}
\newcommand{\ee}{\end{equation}}
\newcommand{\bea}{\begin{eqnarray}}
\newcommand{\eea}{\end{eqnarray}}

\newcommand{\prf}{\noindent {\rm\bf  Proof:\;\; }}

\newcommand{\cA}{{{A}}}

\newcommand{\cT}{{{T}}}

\begin{document}

 \title{How to Resum Feynman Graphs}
\author{Vincent Rivasseau$^a$,\  Zhituo Wang$^b$\\
a) Laboratoire de Physique Th\'eorique, CNRS UMR 8627,\\
Universit\'e Paris XI,  F-91405 Orsay Cedex, France\\
b) Dipartimento di Matematica, Universit\`a di Roma Tre\\
Largo S. L. Murialdo 1, 00146 Roma, Italy\\
E-mail: rivass@th.u-psud.fr, zhituo@mat.uniroma3.it}

\maketitle
\begin{abstract}
In this paper we reformulate the combinatorial core of constructive  quantum field theory. We define
universal rational combinatorial weights for pairs made of a graph and any of its spanning trees. These weights are
simply the percentage of Hepp's sectors of the graph in which the tree is leading, in the sense of Kruskal's greedy algorithm. 
Our main new mathematical result is an integral representation of these weights in term of the positive matrix 
appearing in the symmetric ``BKAR" Taylor forest formula.
Then we explain how the new constructive technique called Loop Vertex Expansion 
reshuffles according to these weights the divergent series of the \emph{intermediate field} representation into
a convergent series which is the Borel sum of the ordinary perturbative Feynman's series.

\end{abstract}

\begin{flushright}
\end{flushright}
\medskip


\medskip

\section{Introduction}

The fundamental step in quantum field theory (QFT) is to compute the logarithm of a functional integral\footnote{The main
feature of QFT is the renormalization group, which is made of a sequence of such fundamental steps, one for each \emph{scale}.}.
The main advantage of the perturbative expansion in QFT  into a sum of Feynman amplitudes is to
perform this computation explicitly: the logarithm of the functional integral
is simply the same sum of Feynman amplitudes restricted to \emph{connected} graphs. The main disadvantage
is that the perturbative series indexed by Feynman graphs typically diverges. Constructive theory
is the right compromise, which allows both to compute logarithms, hence
connected quantities, but through convergent series. However it has the reputation to be
a difficult technical subject.

Perturbative quantum field theory writes quantities of interest (free energies or
connected functions) as sums of amplitudes of connected graphs
\be  S = \sum_{G} A_G . \label{ordinar}
\ee
However such a formula (obtained by expanding in a power series the exponential of the interaction and then illegally commuting the
power series and the functional integral) is \emph{not} a valid definition since
usually, even with cutoffs, even in zero dimension (!) we have
\be  \sum_{G} \vert A_G \vert = \infty .
\ee
This divergence, known since \cite{Dys},
is due to the very large number of graphs of large size. We can say that Feynman graphs {\it  proliferate too fast}.
More precisely the power series in the coupling constant $\lambda$ corresponding to \eqref{ordinar}
has zero radius of convergence\footnote{This can be proved easily for $\phi^4_d$, the Euclidean Bosonic quantum field theory
with quartic interaction in dimension $d$, with fixed ultraviolet cutoff,
where the series behaves as $ \sum_n  (-\lambda)^n K^n n! $. It is expected to remain true also
for the renormalized series without cutoff; this has been proved in the  super-renormalizable cases $d=2,3$ \cite{Jaf,CR}).}.
Nevertheless for the many models built by constructive field theory, the constructive answer is the \emph{Borel sum}
of the perturbative series (see \cite{Riv} and references therein). Hence the perturbative expansion, although divergent, contains all the
information of the theory; but it should be \emph{reshuffled} into a convergent process.

The central basis for the success of constructive theory is that
{\it trees} do {\it not} proliferate as fast as graphs\footnote{This slower proliferation of trees allows
for the local existence theorems in classical mechanics, since classical perturbation theory
is indexed by trees \cite{LinPoinc}.
Hence understanding constructive theory as a recipe to replace Feynman graphs by trees creates also
an interesting  bridge between QFT and classical mechanics.}, and they are sufficient to see connectivity,
hence to compute logarithms. This central fact is not usually emphasized as such in the classical constructive literature \cite{GJ}. It is also
partly obscured by the historic tools which constructive theory borrowed from statistical mechanics, such as lattice cluster and Mayer expansions.

The Loop Vertex Expansion (LVE for short) \cite{R1} is a recent constructive technique to reshuffle the perturbative expansion
into a convergent expansion using canonical combinatorial tools rather than non-canonical lattices.
Initially introduced to analyze \emph{matrix} models with quartic interactions, it has been
extended to arbitrary stable interactions \cite{RW}, shown compatible with
direct space decay estimates \cite{MR1} and with renormalization in simple super-renormalizable cases \cite{Rivasseau:2011df,ZWGW}.
It has also recently been used and improved \cite{Gurau} to organize the $1/N$
expansion \cite{expansion1,expansion2,expansion3} for random \emph{tensors} models \cite{review,universality,uncoloring},
a promising approach to random geometry and quantum gravity in more
than two dimensions \cite{TT1,TT2}.

It is natural to ask how Feynman graphs are regrouped and summed by this LVE.
The purpose of this paper is to answer explicitly this question.
We define a simple but non trivial\footnote{Non-trivial means they are \emph{not} the trivial equally distributed weights
$w(G,T) = 1/ \chi (G)$, where $\chi(G)$, the complexity of $G$, is the number of its spanning trees.} set of positive weights $w(G,T)$,
which we call the \emph{constructive} weights.
These weights are
rational numbers associated to any pair made of a connected graph $G$ and a spanning tree
$T \subset G$, which are normalized so that
\be  \sum_{T \subset G}  w(G,T) =1  \label{bary0} .
\ee
They reduce the essence of constructive theory to the single short equation
\be  S = \sum_{G} A_G = \sum_{G }  \sum_{T \subset G}  w(G,T)  A_G =  \sum_{T} A_T, \quad A_T = \sum_{G \supset T}  w(G,T)  A_G .
\label{const}
\ee
Indeed if we formulate $S$ in terms of the \emph{right graphs}, then
\be  \sum_{T}  \vert A_T\vert < +\infty ,
\label{const1}
\ee
which means that $S$ is now well defined!

In the first section of this paper we define the constructive weights $w(G,T)$ as the percentage of Hepp's sectors \cite{Hepp} of $G$
in which the tree $T$ is leading in the sense of \emph{Kruskal greedy algorithm}  \cite{kruskal}. We then establish an integral representation \eqref{integ} of these weights in terms of the positive-type matrix
which is at the heart of the forest formulas of constructive theory \cite{BK,AR1} and of the LVE \cite{R1}.
Hence this representation connects Hepp's sectors, the essential tools for renormalization in the parametric representation
of Feynman integrals, to the forest formula, the essential tool of the LVE.
It strongly suggests that the LVE should be well-adapted for renormalization, especially
in its parametric representation defined in \cite{Gurau}.

In the second section we explain what are the right graphs to use. In the Bosonic case, they are not the ordinary  Feynman graphs,
but the graphs of the so-called intermediate field representation of the theory. This was the essential discovery of the LVE \cite{R1}.
In the third section we fully explicit up to second order the corresponding graphs and their reshuffling
in the very simple case of the $\phi^4_0$ quantum field theory in zero dimension.
We end up with a conjecture, which, if true, would allow to define QFT in non-integer dimension
of space-time.

\section{The Weights}

\subsection{Paths and Sectors}

We consider from now on pairs $(G,T)$ always made of a \emph{connected} graph $G$ and one of its \emph{spanning} trees $T$.
We denote by $V$ the number of vertices and $E$ the number of edges of $G$.
Graphs with multiple edges and self-loops (called tadpoles in physics)
are definitely allowed, as they occur as Feynman graphs in QFT.

Given such a pair $(G,T)$ and a pair $(i,j)$ of vertices in $G$ there is
a unique \emph{path} $P^T_{ij}$ in $T$ joining $i$ to $j$.
If $\ell $ is an edge of $G \supset T$, we also note
$P^T_\ell$ the unique path  in  $T$ joining the two ends $i$ and $j$ of $\ell$.

A \emph{Hepp sector} $\sigma=\{\sigma(1),\cdots, \sigma(|E|)\}$ of a graph $G$ is an ordering of its edges $E$ \cite{Hepp}, and $|E|$ means the cardinal of the set $E$; hence there are $|E|!$ such sectors.

For any such sector $\sigma \in S(G)$, Kruskal greedy algorithm  \cite{kruskal} defines a particular  tree $T( \sigma)$,
which minimizes $\sum_{\ell \in T}  \sigma (\ell)$ over all  trees of $G$. We call it for short the \emph{leading tree} for $\sigma$.
Let us briefly explain how this works.
The algorithm simply picks the first edge $\ell_1$ in $\sigma$ which is not a self-loop.
The next edge $\ell_2$
in $\sigma$ that does not add a cycle to the (disconnected) graph with vertex set $V$ and edge set $\ell_1$ and so on.
Another way to look at it is through a deletion-contraction recursion: following the ordering of the sector $\sigma$, every edge is either deleted
if it is a self-loop or contracted if it is not. The set of contracted edges is exactly the leading tree for $\sigma$.

Remark that this  leading tree $T(\sigma)$ has been considered intensively in the context of perturbative and constructive
renormalization in QFT \cite{Riv}, as it plays an essential role to get sharp
bounds on renormalized quantitites: it is exactly the leading tree of the Kirchoff-Symanzik polynomial $U_G$ of the parametric representation
(\eqref{sym1}-\eqref{sym2} below) in the Hepp sector $\sigma$.

Remark also that given any sector $\sigma $ the (unordered) tree $T(\sigma)$ comes naturally
equipped with an \emph{induced ordering} (the order in which the edges of $T(\sigma)$
are picked by Kruskal's algorithm). The corresponding ordered tree is noted $\bar T(\sigma)$.

\subsection{Definitions}

There are two equivalent ways to define the constructive weights $w(G,T)$, through paths or through sectors.
The sector definition is simpler as it simply states that $w(G,T)$ is the
percentage of sectors $\sigma$ such that $T(\sigma)=T$.

\begin{definition}
\be w(G,T) =\frac{N(G,T)}{|E|!}  \label{def}
\ee
where  $N(G,T)$ is the number of sectors $\sigma$
such that $T(\sigma)=T$.
\end{definition}

From this definition it is obvious that the $w(G,T)$ form a
probability measure for the spanning trees of a graph, hence that \eqref{bary0}
holds. It is also obvious that these weights are integers divided by $E!$, hence rational numbers.
Remark also that the weights $w(G,T)$ are \emph{symmetric} with respect
to relabeling of the vertices of $T$ (which  are also those of $(G)$). However
the positivity property important for constructive theory is not obvious in this definition.

\begin{theorem}
\be  w(G,T) =  \int_0^1 \prod_{\ell \in T} dw_\ell   \prod_{\ell \not\in T} x^\cT_{\ell}(\{w\})    \label{integ}
\ee
where $x^\cT_{\ell}(\{w\})$ is the minimum over the $w_{\ell'}$ parameters
of the edges $\ell'$ in $P^T_\ell$. If $\ell$ is a self-loop, hence the path is empty, we put $x^\cT_{\ell}(\{w\})=1$.
\end{theorem}

\prf We introduce first parameters $w_\ell$ for all the edges in $G-T$, writing
\be x^\cT_{\ell}(\{w \}) = \int_0^1  dw_\ell \bigl[ \prod_{\ell' \in P^T_\ell} \chi(w_\ell < w_{\ell'} ) \bigr] ,
\ee
where $\chi(\cdots)$ is the characteristic function of the event $\cdots$.
Then we decompose the $w$ integrals according to all possible orderings $\sigma$. We need only prove that
\bea
w(G,T)&=& \int_{0}^1 \prod_{\ell\in G}dw_\ell \prod_{\ell \not\in T} \bigl[ \prod_{\ell' \in P^T_\ell} \chi(w_\ell < w_{\ell'} ) \bigr]
\nonumber \\ &=&
\sum_{\sigma} \chi ( T(\sigma) =T) \int_{0< w_{\sigma(E)}  < \cdots < w_{\sigma(1)}  < 1} \prod_{\ell\in G}dw_\ell   .
\eea
This is true because in the domain of integration defined by $0< w_{\sigma(E)}  < \cdots < w_{\sigma(1)}  < 1$
the function  $\prod_{\ell \not\in T} \bigl[ \prod_{\ell' \in P^T_\ell} \chi(w_\ell < w_{\ell'} ) \bigr]$
is zero or 1 depending whether $ T(\sigma) =T$ or not, as this function being 1 is
exactly the condition for Kruskal's algorithm to pick exactly $T$. Strict inequalities are easier to use here: of course
equal values of $w$ factors have zero measure anyway. Hence
\be  \int_0^1 \prod_{\ell \in T} dw_\ell   \prod_{\ell \not\in T} x^\cT_{\ell}(\{w\})  =\frac{N(G,T)}{|E|!}  \; .  \label{def3}
\ee
\qed

This theorem provides an integral representation of the weights, in terms of  ``weakening parameters"
$w_\ell$ for the edges $\ell \in T$.
The fundamental advantage of the constructive weights $w(G,T)$ over naive uniform weights
is precisely the positivity property of the  $x^\cT_{\ell}(\{w\})$
matrix, which we now explain.

\subsection{Positivity}

To any triple $(G,T, \sigma)$ is associated a sequence of $V$
partitions $B_k$, $k = 1, \cdots , V$, of the set of vertices of $G$ into disjoint blocks, which are the connected components of the
sequence of forests obtained when constructing the ordered tree $\bar T(\sigma)$. More precisely
the first partition $B_1$ is made of singletons, one for each vertex of $V$; the second partition is made of the connected components
of the forest $F_1$ made of the first edge of $T(\sigma)$, and so on until $B^V$ which is made of a single connected component
containing all vertices of $G$. Clearly there are exactly $V-i+1$ disjoint blocks in $B_i$, labeled as $B_k^a$, $a = 1, \cdots V-i+1$.
Remark that these partitions only depend on $\bar \sigma $, the restriction
of the ordering $\sigma$ to $T$.

The $V$ by $V$ real symmetric block matrix $B_k(T,\bar \sigma)_{ij}$
with 1 between elements $i,j$ belonging to the same connected component  $B_k^a$ and 0
between elements $i,j$ belonging to different connected component  $B^k_a$ at stage $k$
is obviously positive (although not positive definite as soon as  blocks are not trivial).

\begin{theorem}[Positivity] Let us define the $V$ by $V$ real symmetric matrix
 $x^\cT_{ij}(\{w\})$ as in Theorem 2.1, that is with 1 on the diagonal $i=j$ and as the
minimum over the $w_{\ell'}$ parameters
over the lines $\ell'$ in $P^T_{ij}$ for $i \ne j$. This matrix
is \emph{positive semidefinite} for any $w_\ell \in [0,1]^{V-1}$. It is \emph{positive definite} for any $w_\ell \in [0,1[^{V-1}$.
\end{theorem}

\prf This is the central property of the forest formula \cite{BK,AR1}. We recall briefly the proof for completeness.
Consider a fixed value of the $w_\ell \in [0,1]^{V-1}$. There is at least one
sector $\bar \sigma$ of $T$ to which it belongs, hence such that
\be  0 \equiv w_{\bar\sigma(V)} \le   w_{\bar\sigma(V-1)}  \le \cdots  \le w_{\bar\sigma(k)} \le \cdots  u_{\bar\sigma(1)}  \le 1 \equiv u_{\bar\sigma(0)}
\ee
We have then the  decomposition
\be  x^\cT_{ij}(\{w\}) = \sum_{k=1}^V  \bigl[ w_{\bar\sigma(k-1)}   - w_{\bar\sigma (k)} \bigr] B_k(T,\bar \sigma)_{ij} .
\ee
which proves that $x^\cT_{ij}(\{w\})$, as a barycenter of positive type matrices with positives weights, is positive type. Furthermore for
$w_\ell \in [0,1[^{V-1}$, the coefficient of the identity in this barycentric decomposition is non zero, hence the matrix $x^\cT_{ij}(\{w\}) $
is positive definite in that case.
\qed

\subsection{Example}

\begin{figure}[!htb]
\centering
\includegraphics[scale=0.8]{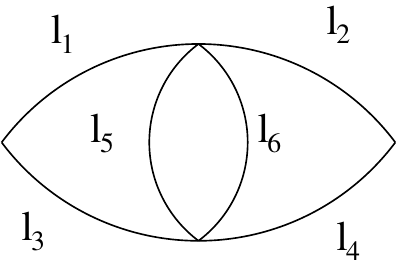}
\caption{The graph G with 6 edges and 12 spanning trees.}
\label{eyea}
\end{figure}

Let us consider the graph $G$ of  Fig. \ref{eyea}. It has 6 edges
$\{l_1, l_2, l_3, l_4, l_5, l_6\}$ and 12 spanning trees:
\begin{eqnarray}
  \{l_1, l_2, l_3\},  \{l_1, l_2, l_4\},  \{l_1, l_3, l_4\},  \{l_2, l_3, l_4\},  \{l_1, l_2, l_5\},\{l_1, l_2, l_6\},\nonumber\\  \{l_3, l_4, l_5\},
\{l_3, l_4, l_6\},  \{l_1, l_4, l_5\},  \{l_1, l_4, l_6\},  \{l_2, l_3, l_5\},  \{l_2, l_3, l_6\}.
\end{eqnarray}

\begin{figure}[!htb]
\centering
\includegraphics[scale=0.8]{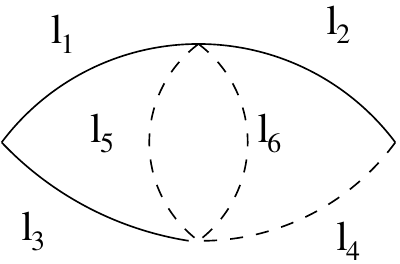}
\caption{The spanning tree $\cT_{123}$. The dotted edges add cycles to the tree.}
\label{eyeb}
\end{figure}

Let us compute the constructive weights $w(G,T)$ for each of these trees. To each edge $l_i$ we associate a factor $w_i$.
Consider first the spanning tree $\cT_{123}=\{l_1, l_2, l_3\}$ , see Figure(\ref{eyeb}). The edges not in the tree are $l_4$, $l_5$ and $l_6$.
The weakening factor for $l_5$ and $l_6$ is $\inf(w_1, w_3)$ and
the weakening factor for $l_4$ is $\inf(w_1, w_2, w_3)$. Therefore we have
\begin{eqnarray}
&& w(G,\cT_{123})=\int_{0}^1 \int_{0}^1 \int_{0}^1dw_1 dw_2 dw_3\inf(w_1, w_3)^2\inf(w_1, w_2, w_3)
\end{eqnarray}
We compute only two of the integrals explicitly as others are obtained
by changing the names of variables.
\begin{equation}
\int_{w_1< w_2< w_3}dw_1 dw_2 dw_3\  w_1^3=\int_0^1 dw_3\int_0^{w_3} dw_2 \int_0^{w_2}dw_1 w_1^3=  \frac{1}{120},
\end{equation}
\begin{equation}
 \int_{w_2< w_1< w_3}dw_1 dw_2 dw_3\  w_3^2\  w_2=\frac{1}{60} .
\end{equation}
So we have
\begin{equation}
w(G,\cT_{123})=\frac{1}{120}\times 4+\frac{1}{60}\times 2=\frac{1}{15}.
\end{equation}
The constructive weights in $G$ of the spanning trees $\cT_{124}$, $\cT_{134}$ and $\cT_{234}$ are the same.

\begin{figure}[!htb]
\centering
\includegraphics[scale=0.8]{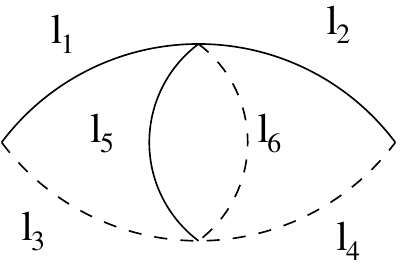}
\caption{The spanning tree $\{l_1, l_2, l_5\}$}
\label{eyec}
\end{figure}

Next we consider the tree $\{l_1, l_2, l_5\}$. (See Figure \ref{eyec}).
The weakening factors are $\inf(w_1, w_5)$ for loop line $l_3 $, $\inf(w_2, w_5)$ for loop line $l_4$ and $w_5$
for loop line $l_6$. Hence one finds
\begin{eqnarray}
&& w(G,\cT_{125})=\int_{0}^1 \int_{0}^1 \int_{0}^1dw_1 dw_2 dw_5\inf(w_1, w_5)\inf( w_2, w_5) w_5\end{eqnarray}
We have
\begin{equation}
 \int_{w_1< w_2< w_5}dw_1 dw_2 dw_5 w_1 w_2 w_5=\frac{1}{48},
\end{equation}
\begin{equation}
 \int_{w_5< w_1< w_2}dw_1 dw_2 dw_5 w_5^3=\frac{1}{120},
\end{equation}
\begin{equation}
 \int_{w_2< w_5< w_1}dw_1 dw_2 dw_5  w_2 w^2_5=\frac{1}{60}.
\end{equation}
Hence
\begin{equation}
 w(G,\cT_{125})=\frac{1}{120}\times 2+\frac{1}{60}\times 2 +\frac{1}{48}\times 2=\frac{11}{120} .
\end{equation}
This is also the constructive weight of trees
$\cT_{126}, \cT_{345}, \cT_{346}, \cT_{125}, \cT_{145}, \cT_{146}, \cT_{235}$ and $\cT_{236}$.

We can check that
\begin{equation}
 \sum_{\cT \in G} w(G, \cT) = 4.\frac{1}{15}+ 8.\frac{11}{120} =1.
\end{equation}
We remark that 6! = 720, hence that $N(G',\cT_{123})=48$ and  $N(G',\cT_{125})= 66$. This
can be checked by direct counting of the sectors $\sigma$ with $T(\sigma) =\cT_{123}$ or $T(\sigma) =\cT_{125}$.
The 48 sectors with $T(\sigma) =\cT_{123}$ are the thirty-six sectors with $\{1,2,3\}$ being the set of the first three edges,
plus the six sectors 135624, 136524, 135264, 135246, 136254, 136245 and the six analogs with
1 and 3 exchanged. The 66 sectors with $T(\sigma) =\cT_{125}$ are the 36 with $\{1,2,5\}$ being the set of the first three edges,
plus 30 others: six starting with 15 with third edge either 3 or 6; six analogs starting with 25
with third edge either 4 or 6; 6 starting with 52 with third edge either 4 or 6,   6 analogs  starting with 15
with third edge either 3 or 6, and finally six sectors starting with 56 with third edge either 1 or 2.

\section{The Graphs}

\subsection{Naive Repacking}

Consider the expansion (\ref{ordinar}) of a connected quantity $S$.
Reordering ordinary Feynman perturbation theory according to trees with relation
\eqref{const} rearranges the Feynman expansion according to trees
with the same number of vertices as the initial graph. Hence it reshuffles
the various terms of a {\it given, fixed} order of perturbation theory. Remark
that if the initial graphs have say degree 4 at each vertex, only
trees with degree less than or equal to 4 occur in the rearranged
tree expansion.

For Fermionic theories this is typically sufficient and one has for small enough coupling
\be  \sum_{\cT}  \vert \cA_\cT \vert < \infty
\ee
because Fermionic graphs mostly compensate each other
at a fixed order by Pauli's principle; mathematically this is because these
graphs form a determinant and the size of a determinant is much less
than what its permutation expansion suggests. This is well known
\cite{Les,FMRT1,AR2}.

But this naive repacking fails for Bosonic theories, because we know the graphs at given order add up with the same sign!
Hence the only interesting reshuffling must occur between graphs of different orders.
\subsection{The Loop Vertex Expansion}

The initial formulation of the loop vertex expansion \cite{R1}
consists in applying the forest formula of \cite{BK,AR1} to the intermediate field representation.
As we explain now, it can also be reformulated as \eqref{const}
but for the graphs of this intermediate field
representation, which resums an infinite number of pieces of the ordinary graphs.

Recall first that since the combinatorics of Feynman graphs requires labeling the half-edges or fields $\phi$
at every vertex of coordination $d$ as $\phi_1 ,  \cdots \phi_d$, each Feynman vertex is in fact equipped with
a \emph{ciliated cyclic ordering} of its edges. The cilium gives a starting point
and the cyclic ordering allows to then label all fields from this starting point.
This is the reason for which Feynman graphs below are represented as ribbon graphs.

The principle of the intermediate field representation is to decompose any interaction of degree higher than three
in terms of simpler three-body interactions. It is an extremely useful idea, with deep applications
both to mathematics and physics.
Quantum field theory, in particular, often discovered an intermediate field and its corresponding physical particles
inside what was initially considered as local four body interactions\footnote{Recall
that intermediate field representations are particularly natural for 4-body interactions
but can be generalized to higher interactions as well \cite{RW}.}.

It is easy to describe the intermediate field method in terms of functional integrals, as it is a simple generalization of the formula
\be  e^{- \lambda \phi^4/2}  = \frac{1}{\sqrt{2\pi}}\int e^{- \sigma^2 /2} e^{i \sqrt \lambda \sigma \phi^2} d\sigma .
\ee
In this section we introduce the graphical procedure equivalent to this formula for the simple case of the
 $\phi^4$ interaction.

In that case each vertex has exactly four half-lines.
There are exactly three ways to pair these half-lines into two pairs. Hence
each fully labeled (vacuum) graph of order $n$ (with labels on vertices and half-lines),
which has $2n$ lines
can be decomposed exactly into $3^n$ labeled graphs $G'$
with degree 3 and two different types of lines
\begin{itemize}
\item the $2n$ old ordinary lines

\item $n$ new dotted lines which indicate the pairing chosen at each vertex (see Figure 5).
\end{itemize}

Such graphs $G'$ are called the  3-body extensions of $G$ and we write
$G' {\ \rm ext\ } G$ when $G'$ is an extension of $G$. Let us introduce
for each such extension $G'$ an amplitude $A_{G'} = 3^{-n} A_G$ so that

\be  A_G =  \sum_{G' {\ \rm ext\ } G} A_{G'}
\ee
when $G'$ is an extension of $G$.

Now the ordinary lines of any extension $G'$ of any $G$
must form cycles. These cycles are joined by dotted lines.

\begin{figure}[!htb]
\centering
\includegraphics[scale=0.6]{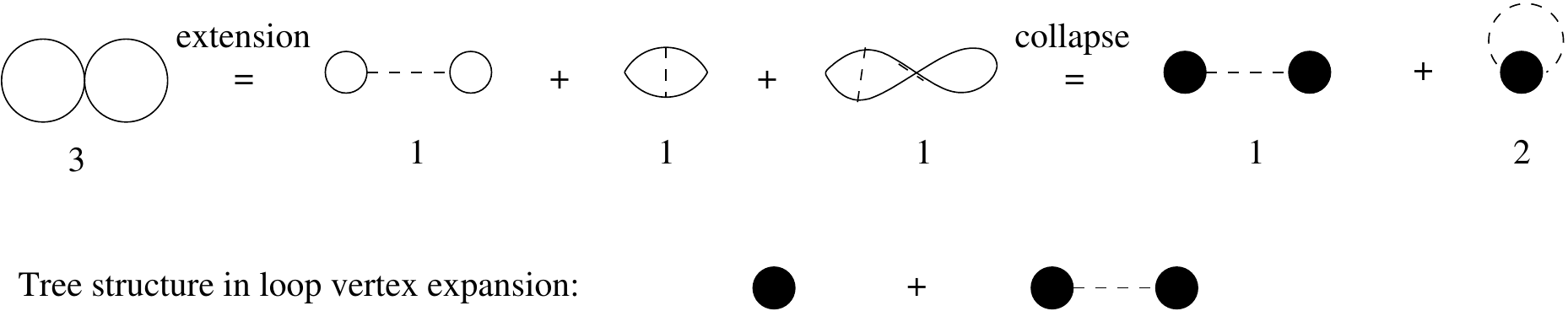}
\caption{The extension and collapse for order 1 graph, with combinatorial weights shown below. The symbol "=" means that the amplitudes of extended graphs and collapsed graphs are the same as those of the initial Feynman graph; only the combinatorial weight in front is reshuffled to attribute Wick contractions
to different drawings.
}
\label{odonea}
\end{figure}

\begin{definition}\label{ctree}
We define the collapse $\bar G'$ of such a graph $G'$ as
the graph obtained by contracting each cycle to a "bold" vertex (see Figure \ref{odonea}).
We write $\bar G' {\ \rm coll\ } G'$ if $\bar G'$ is the collapse of $G'$, and define
the amplitude of the collapsed graph $ \bar G'$ as equal to that of $G'$, which is equal to the amplitude of $G$. And $\bar T$ is defined as the spanning tree
of the collapsed graph $\bar G'$.
\end{definition}

Remark that collapsed graphs, made of bold vertices and dotted lines,
can have now arbitrary degree at each vertex.
Remark also that several different extensions of a graph $G$
can have different collapsed graphs, see Figure \ref{odonea}.

The loop vertex expansion rewrites
\be  S = \sum_G A_G  = \sum_{G' {\ \rm ext\ } G} A_{G'} =
 \sum_{\bar G' {\ \rm coll\ } G' {\ \rm ext\ } G}   A_{\bar G'} .
\ee
Now we perform the tree repacking according to the graphs $\bar G'$
with the $n$ dotted lines and {\it not} with respect to $G$. This is a completely
different repacking:
\be  A_{\bar G'} = \sum_{\bar \cT \subset \bar G'} w(\bar G', \bar \cT) A_{\bar G'},
\ee
so that
\be  S= \sum_{G' {\ \rm ext\ } G}   A_{\bar G'}  = \sum_{\bar \cT \subset \bar G'} A_{\bar \cT},
\ee
\be
A_{\bar \cT} = {\Large \cal B} \biggl( \sum_{\bar G' \supset \bar \cT} w(\bar G',\bar \cT) A_{\bar G'} \biggr) \label{repacked}.
\ee
In equation \eqref{repacked} the left-hand side is defined by the LVE (as a functional integral
over a certain interpolated Gaussian measure for intermediate fields associated to the vertices
of $T$). The meaning of the symbol ${\Large \cal B}$ (where $B$ stands for ``Borel") in \eqref{repacked} is that this left-hand side, as function of the coupling constant of the theory,
is the \emph{Borel sum} of the infinite (divergent) series in the right hand-side.
The main advantage of this repacking over the initial perturbative expansion is:

\begin{theorem} For $\lambda$ small
\be  \sum_{\bar \cT}  \vert A_{\bar \cT} \vert < \infty
\ee
the result being the {\it Borel sum} of the initial perturbative series.
\end{theorem}

The proof of the theorem will not be recalled here
(see  \cite{R1,MR1,R2}) but it relies on the positivity property of
the $ x^\cT_{\ell}(\{w\})$ symmetric matrix, and the representation of each
$A_{\bar \cT} $ amplitude as an integral over a corresponding normalized
Gaussian measure of a product of resolvents bounded by 1.  This convergence
would not be true if we had chosen naive $w(\cT,G)$ equally distributed weights.

 \section{Examples of extensions and collapses}
In this section we give the extension and collapse of the Feynman graphs for $Z$ and $\log Z$ for the $\phi^4_0$ model up to order 2.
We also recover the combinatorics of those graphs
through the ordinary functional integral formula
for the loop vertex expansion formula of \cite{R2}.

The extension and collapse at order 1 was shown in Figure \ref{odonea}. In this case the tree structure is easy.
We find only the trivial "empty" tree with one vertex and no edge and
the "almost trivial"  tree with two vertices and a single edge. The weight for these trees is 1.

At second order we find one disconnected Feynman graph and two connected ones. Only the
connected ones survive in the expansion of $\log Z$.
\begin{figure}[!htb]
\centering
\includegraphics[scale=0.55]{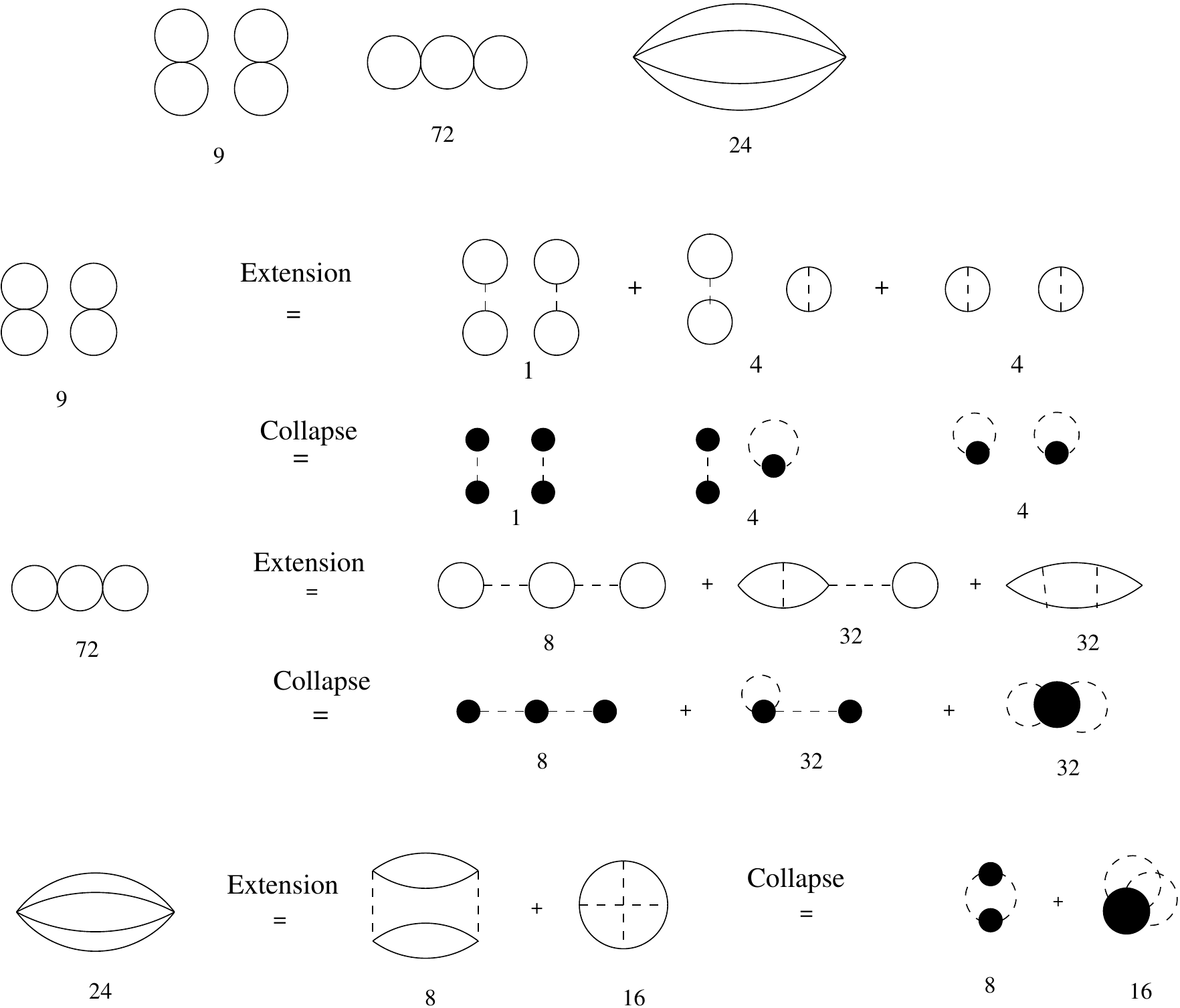}
\caption{The extension and collapse for order 2 graph and their combinatorial factors.}
\label{odtwo}
\end{figure}

\begin{figure}[!htb]
\centering
\includegraphics[scale=0.6]{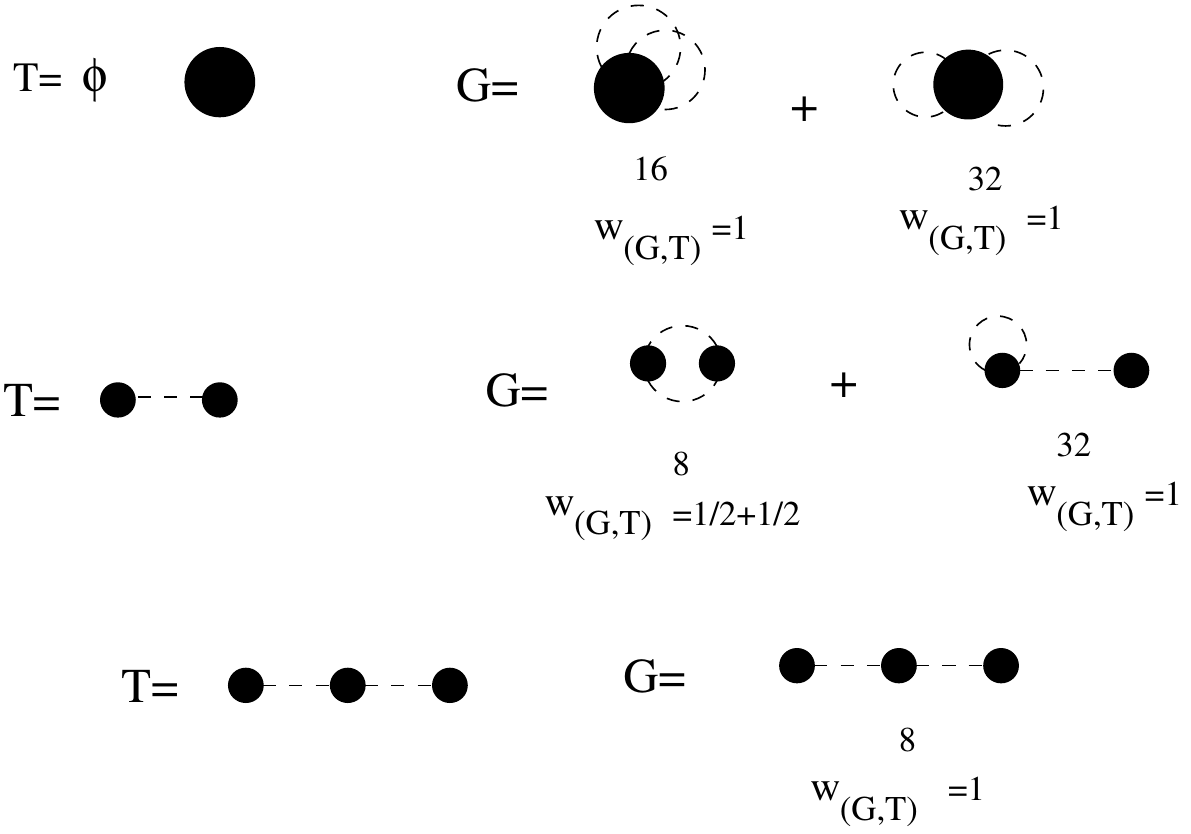}
\caption{The connected graphs and the tree structure from the loop vertex expansion. Remark that Feynman graphs of different orders may have the same tree structure in the LVE representation.}
\label{logtree}
\end{figure}

The corresponding graphs and tree structures are shown in Figure \ref{odtwo} and \ref{logtree}. Using the loop vertex expansion formula we begin to see that graphs
coming from different orders of the expansion of $\lambda$ can be associated to the same tree by the loop vertex expansion. Indeed we recover contributions for the trivial and
almost trivial trees of the previous figure. But we find also a new contribution belonging to a tree with two edges.

From these examples we find that the structure of  the loop vertex expansion is totally different from that of Feynman graph calculus. At each order of the loop vertex expansion it combines terms in different orders of $\lambda$.

\section{Non-integer Dimension}

Let us now consider, e.g. for $0< D \le 2$ the Feynman amplitudes for the $\phi^4_D$
theory.
They are given by the following convergent parametric representation (see e.g. \cite{tutte} for a recent reference)
\be  A_{D,G} = \int _0^\infty d\alpha \frac{e^{-m^2\sum_\ell \alpha_\ell}  }{U_G^{D/2}}
\label{sym1}
\ee
where $m$ is the mass and $U_G$ is the Kirchoff-Symanzik polynomial for $G$
\be  U_G  = \sum_{\cT \in G}  \prod_{\ell \not \in \cT}  \alpha_\ell  . \label{sym2}
\ee

All the previous decompositions working at the level of graphs, they are independent
of the space-time dimension.
We know that for $D=0$ and $D=1$ the loop vertex expansion is convergent.
Therefore it is tempting to conjecture, for instance at least for $D$ real and $0 \le D <2$
(that is when no ultraviolet divergences require renormalization),
that repacking in the same way the series of Feynman amplitudes
in non-integer dimension also works, that is, after introducing the same extensions and collapse operations:

\begin{conjecture} The series $ \sum_{\bar G' \supset \bar \cT}  w( \bar G' , \bar \cT ) A_{D, \bar G'}  $
is Borel summable in the coupling constant of the theory for any real $D$ with $0 \le D <2$ and denoting
$A_{D, \bar \cT} $ its Borel sum:
\be   A_{D, \bar \cT}  =  {\Large \cal B} \biggl(  \sum_{\bar G' \supset  \bar \cT}  w(\bar G', \bar \cT) A_{D,\bar G'}  \biggr),
\ee
the series $\sum_{\bar \cT}  A_{D, \bar \cT}$ is absolutely convergent  for $\lambda$ small:
\be  \sum_{\bar \cT}  \vert A_{D, \bar \cT} \vert < \infty \; ,
\ee
the result being the Borel sum of the initial perturbation series.
\end{conjecture}

If true this conjecture would allow rigorous interpolation
between quantum field theories in various dimensions of space time. It could e.g. lead to a possible
justification of the Wilson-Fisher $4-\epsilon$ expansion that allows good numerical approximate
computations of critical indices in 3 dimensions.

An other approach  to quantum field theory in non integer dimension,
also based on the forest formula but  more radical,
is proposed in \cite{GMR}.

\medskip
\noindent{\bf Acknowledgments}
We thank H.  Kn\"orrer for asking the question which lead to writing this paper
and R. Gurau for interesting discussions.  The research has been partly supported by the European Research Council under the European Union's Seventh Framework Programme ERC Starting Grant CoMBoS (grant agreement n.239694).

\end{document}